\documentclass[letter, scriptaddress, twocolumn,  prd, showpacs,showkeys]{revtex4-1}

	\usepackage{amsmath}
        \usepackage{mathrsfs}
	\usepackage{amssymb}
	\usepackage{graphicx} 
	\usepackage{makeidx}
	\usepackage{amsfonts}
	\usepackage[ansinew]{inputenc}
	\usepackage[usenames, dvipsnames]{pstricks}
	\usepackage{epsfig}
	\usepackage{pst-grad} % For gradients
	\usepackage{pst-plot} % For axes
	\usepackage[colorlinks, hyperindex]{hyperref}
	\hypersetup
	{
		colorlinks, %
		citecolor=blue, %
		linkcolor=Green, %
		urlcolor=blue, %
	}

\makeindex

\begin{document}

\title{{\Large\bf Addressing the issue of non-unitarity in anisotropic quantum cosmology}}

\author{Sridip Pal} 
\email{sridippaliiser@gmail.com}
\author{Narayan Banerjee}
\email{narayan@iiserkol.ac.in}
\affiliation{Department of Physical Sciences, \\
Indian Institute of Science Education and Research - Kolkata,\\
Mohanpur Campus, District Nadia,\\ 
West Bengal 741252, India.}

\begin{abstract}
In the present work we show that the widely believed pathology of the non-unitarity of anisotropic quantum cosmological models cannot be a generic problem. We exhibit a non trivial example, a Bianchi-I model with an ultrarelativistic fluid, that has a well behaved time independent norm. We also show that a suitable operator ordering should produce time independent norms for the wave packets in the case of other more realistic fluids as well. 
\end{abstract}
\maketitle
In the absence of a generally accepted quantum theory of general relativity, quantum mechanical principles are applied to many individual gravitational systems. Cosmological models are certainly amongst the fields where this kind of quantization finds an application. The universe at its early stage of evolution, at an energy scale where classical general relativity loses its viability, requires a quantum description. For some excellent reviews, we refer to \cite{wilt, halli}. Quantum Cosmology has, however, many issues yet to be resolved. As time is a coordinate in a relativistic theory of gravity, there is a problem of the identification of a suitable time parameter against which the evolution of the universe would be described \cite{kuchar1, isham, rovelli, anderson}. Moreover, the interpretation of the wave function faces a challenge in quantum cosmology. The Copenhagen interpretation fails as there is no exterior observer for the system. There are attempts in this directi
 on with a many 
 world interpretation and with Bohmian trajectories \cite{nelson1}. Problems regarding the imposition of proper boundary conditions are there as well \cite{wilt}.  \\

The present work deals with another widely known problem, the alleged non-unitarity of the anisotropic quantum cosmological models. The corresponding Hamiltonian, although hermitian,  is not self-adjoint. The norm of the wave packet is thus time dependent and hence there is a non-conservation of probability. It may be argued that the observed universe is isotropic but one is not really certain  whether the very early universe, beyond the Planck scale, is actually isotropic or not! Furthermore, this feature definitely makes the scheme of quantization unreliable. For a very recent review, we refer to the work of Pinto-Neto and Fabris \cite{nelson1}.   \\

There is a scheme of quantization of a cosmological model with a matter field, namely a perfect fluid. Following Schutz's formalism, where the fluid variables are given dynamical degrees of freedom \cite{schutz1, schutz2}, the relevant action can be written in terms of the metric tensor components representing the gravity sector and some thermodynamic potentials representing the fluid sector. The method had been used by Lapchinskii and Rubakov \cite{rubakov} for a Friedmann model.  Recently the method has been utilized by Alvarenga and Lemos \cite{alvarenga1}, Batista et al \cite{batista}, Alvarenga et al \cite{alvarenga2}, Vakili \cite{vakili1, vakili2}, Alvarenga et al\cite{alvarenga3}, Majumder and Banerjee \cite{barun}. The latter two, dealing with anisotropic Bianchi models, show that the models are non-unitary. If no proper matter field is used, resulting in no physical identification of ``time'', this problem of non-conservation of probability may remain unnoticed\cite{lidsey, nelson2}.   \\ 

In what follows, we show that this general belief is actually not quite correct. We take up a Bianchi I cosmological model with a perfect fluid as an example. Using Schutz's formalism, we quantize this model for a barotropic equation of state $P=\alpha \rho$ where $P$ and $\rho$ are the pressure and density of the perfect fluid and $\alpha$ is a constant where $\alpha \leq 1$. \\

The relevant action is given by
\begin{equation}\label{Action}
\mathcal{A}=\int_{M}d^{4}x \sqrt{-g}R +2 \int_{\partial M} \sqrt{h}h_{ab}K^{ab}+\int_{M} d^{4}x \sqrt{-g}P,
\end{equation}
where $K^{ab}$ is the extrinsic curvature, and $h^{ab}$ is the induced metric over the boundary $\partial M$  of the 4 dimensional space-time manifold $M$. The units are so chosen that $16\pi G = 1.$ The metric of Bianchi type-I model is given by
\begin{equation}\label{Metric}
ds^{2}=n^{2}dt^{2}-\left[a^{2}(t)dx^{2}+b^{2}(t)dy^{2}+c^{2}(t)dz^{2}\right],
\end{equation}
where $n(t)$ is the lapse function and $a,b,c$ are functions of the cosmic time $t$. 

Using the metric we rewrite the gravity sector of \eqref{Action} in the following form
\begin{equation}\label{GRaction}
\mathcal{A}_g =\int dt \left[-\frac{2}{n}\left(\dot{a}\dot{b}c+\dot{b}\dot{c}a+\dot{c}\dot{a}b\right)\right].
\end{equation}
We introduce a set of new variables as
\begin{eqnarray}\label{coordinate}
a(t)&=&e^{\beta_{0}+\beta_{+}+\sqrt{3}\beta_{-}},\\
b(t)&=&e^{\beta_{0}+\beta_{+}-\sqrt{3}\beta_{-}},\\
c(t)&=&e^{\beta_{0}-2\beta_{+}}.
\end{eqnarray}
This choice of variables, $\beta_0, \beta_+, \beta_-$ are not new\cite{alvarenga3, barun}. The Lagrangian density of the gravity sector now becomes
\begin{equation}\label{7}
\mathcal{L}_{g}=-6\frac{e^{3\beta_{0}}}{n}\left(\dot{\beta}_{0}^{2}-\dot{\beta}_{-}^{2}-\dot{\beta}_{+}^{2}\right).
\end{equation}
The conjugate momenta, as a consequence, are given by
\begin{eqnarray}\label{8}
p_{0}=-12\frac{e^{3\beta_{0}}}{n}\dot{\beta}_{0},\\
p_{\pm}=12\frac{e^{3\beta_{0}}}{n}\dot{\beta}_{\pm}.
\end{eqnarray}
The corresponding Hamiltonian becomes
\begin{equation}\label{hamgrav}
 H_{g}=-n\exp(-3\beta_{0})\left\{\frac{1}{24}\left(p_{0}^{2}-p_{+}^{2}-p_{-}^{2}\right)\right\}.
\end{equation}

In Schutz's formalism \cite{schutz1, schutz2} the fluid velocity $U_{\nu}$ is given by
\begin{equation}
\label{3.1}
U_{\nu}=\frac{1}{h}\left(\partial_{\nu}\epsilon+\theta\partial_{\nu}S\right),
\end{equation}
for a spacetime without any vorticity. Here $S$ is specific entropy and $h$ is specific enthalpy. The potentials $\epsilon$ and $\theta$ do not have any direct physical significance. The velocity is normalised as $U_{\nu}U^{\nu}=1.$

Using standard thermodynamical considerations, the fluid part of the action \eqref{Action} can be cast into the form:
\begin{equation}
\label{3.91}
\begin{split}
\mathcal{A}_{f}&=\int dt \mathcal{L}_{f}\\&= V \int dt \left[n^{-\frac{1}{\alpha}}e^{3\beta_{0}}\frac{\alpha}{\left(1+\alpha\right)^{1+\frac{1}{\alpha}}}\left(\dot{\epsilon}+\theta\dot{S}\right)^{1+\frac{1}{\alpha}}e^{-\frac{S}{\alpha}}\right],
\end{split}
\end{equation}
where the factor of V comes out due to space integration as we are dealing with a homogeneous model.

We define canonical momentum to be $p_{\epsilon}=\frac{\partial\mathcal{L}_{f}}{\partial\dot{\epsilon}}$ and
$p_{S}=\frac{\partial\mathcal{L}_{f}}{\partial\dot{S}}$ and Hamiltonian becomes 
\begin{equation}
\label{hamfl}
H_{f}=ne^{-3\alpha\beta_{0}}p_{\epsilon}^{\alpha +1}e^{S}.
\end{equation}
We effect the canonical transformation,
\begin{eqnarray}\label{9}
T&=&-p_{S}\exp(-S)p_{\epsilon}^{-\alpha -1},\\
p_{T}&=&p_{\epsilon}^{\alpha+1}\exp(S),\\
\epsilon^{\prime}&=&\epsilon+\left(\alpha+1\right)\frac{p_{S}}{p_{\epsilon}},\\
p_{\epsilon}^{\prime}&=&p_{\epsilon},
\end{eqnarray}

so that the Hamiltonian \eqref{hamfl} of the matter sector becomes
\begin{equation}
\label{3.93}
H_{f}= ne^{-3\beta_{0}}e^{3\left(1-\alpha\right)\beta_{0}}p_{T}.
\end{equation}
The Poisson brackets $\left\{\epsilon^{\prime},p_{\epsilon}^{\prime}\right\}=1$ and $\left\{T,p_{T}\right\}=1$ are satisfied with all other Poisson brackets being $0$. This ensures the canonical structure of the new variables. \\

For a canonical quantization, the Poisson bracket is replaced by a commutator bracket, and we have 
\begin{equation}
\label{3.95}
\left[T,-\imath \frac{\partial}{\partial T}\right]=\imath.
\end{equation}
In Schrodinger equation, time derivative appears in the first order so as to guarantee positive normed state. Here the fluid momentum, which is conjugate to newly defined variable $T$, given by  $-\imath \frac{\partial}{\partial T}$ as an operator, comes as a linear term in Hamiltonian. It is thus a simple indication that one can pick up $T$ as the time parameter. So we have first derivative with respect to time as required for the positivity of norm. The second justification, which is necessary so as to give time an orientation and the same direction as cosmic time, comes classically from the equation
\begin{equation}
\frac{1}{n}\frac{dT}{dt}=\left\{T,H_{f}\right\}= e^{-3\alpha\beta_{0}} > 0.
\end{equation}
The fact $\frac{1}{n}\frac{dT}{dt}$ has the same sign everywhere makes it orientable while its positivity gives $T$ the same arrow as the cosmic time $t$.  \\

The net Hamiltonian for the gravity plus the matter sector now becomes
\begin{equation}\label{10}
H=n\exp(-3\beta_{0})\left\{-\frac{1}{24}\left(p_{0}^{2}-p_{+}^{2}-p_{-}^{2}\right)+e^{3(1-\alpha)\beta_{0}}p_{T}\right\}.
\end{equation}
Variation of the action, with respect to $n$, yields Hamiltonian constraint
\begin{equation}\label{11}
\mathcal{H}=\frac{1}{n}H=0.
\end{equation}

We now promote the super Hamiltonian $\mathcal{H}$ to an operator and postulate commutation relation amongst the quantum operators as usual. Unlike the fluid momentum, the momenta coming from the pure gravity sector come as quadratic terms, so they define kinetic energy in the Hamiltonian. \\

We write $p_{i}\mapsto -\imath\hbar\partial_{\beta_{i}}$ for $i=0,+,-$, and $p_{T} \mapsto -\imath \hbar\partial_{T}.$
This mapping is equivalent to postulating the fundamental commutation relations
\begin{equation}\label{13.1}
\left[\beta_{i},p_{j}\right]=\imath\hbar\delta_{ij}\mathbb{I}.
\end{equation}

Henceforth we shall use natural units, i.e., $\hbar = 1$. It deserves mention that a simple choice of gauge as $n=e^{3\alpha{\beta}_{0}}$ will make $T$ independent of ${\beta}_{0}$ and $\frac{\partial}{\partial T}$ then commutes with the gravity sector. This provides the justification of using ${\beta}_{i}$ and $T$ as a set of independent coordinates. The Wheeler-De Witt equation, $\mathcal{H}\psi =0,$ now takes the form 
\begin{equation}\label{WDeq}
\left(\frac{\partial^{2}}{\partial \beta^{2}_{0}}-\frac{\partial^{2}}{\partial\beta^{2}_{+}}-\frac{\partial^{2}}{\partial\beta^{2}_{-}}\right)\psi = 24\imath e^{3(1-\alpha)\beta_{0}}\frac{\partial \psi}{\partial T}.
\end{equation} 

We assume that the wave function $\psi$ is separable as $\psi(\beta_{0},\beta_{+},\beta_{-},T)=\phi(\beta_{0},\beta_{+},\beta_{-})e^{-\imath ET}$ where $E$ is a constant.\\

So the equation \eqref{WDeq} becomes
\begin{equation}\label{17}
\left(\frac{\partial^{2}}{\partial \beta^{2}_{0}}-\frac{\partial^{2}}{\partial\beta^{2}_{+}}-\frac{\partial^{2}}{\partial\beta^{2}_{-}}\right)\phi =24E\phi e^{3(1-\alpha)\beta_{0}}.
\end{equation}

The solution for this equation is now discussed for two different cases, namely $\alpha = 1$ and $\alpha \neq 1$.\\

{\bf Case I: Stiff fluid ($\rho=P$):}\\

We put $\alpha=1$, i.e., $P=\rho$, and again use a separation of variables as $\phi=\phi_{0}(\beta_{0})\phi_{+}(\beta_{+})\phi_{-}(\beta_{-})$ to obtain $\phi_{\pm}=C_{\pm}e^{\imath k_{\pm}\beta_{\pm}}$
and as a result, the behaviour of $\phi_{0}$ is determined by
\begin{equation}\label{phi0}
\frac{\partial^{2}\phi_{0}}{\partial \beta^{2}_{0}}+\left(-24E+\left(k_{+}^{2}+k_{-}^{2}\right)\right)\phi_{0}=0.
\end{equation}

Hence the general solution can be written as
\begin{equation}\label{20}
\psi = e^{\imath k_{+}\beta_{+}}e^{\imath k_{-}\beta_{-}}e^{\imath \omega \beta_{0}}e^{-\imath ET}
\end{equation}
where $\omega^{2}=-24E+k_{+}^{2}+k_{-}^{2}.$ Initially, we have $k_{\pm}$ and $E$ as free parameters and $\omega$ is defined as a function of $k_{\pm}$ and $E$. Once we have a relation between 4 variables $\omega$, $k_{\pm}$ and $E$, we can mathematically treat any three of them as independent variables while the 4th one will be a function of other three. Here we treat $k_{\pm}$ and $\omega$ as independent variables while $E$ has to be understood as a function of $k_{\pm}$ and $\omega$ where $k_{\pm}\in \left(-\infty,\infty\right)$.

One can thus construct the following wave packet via superposing solutions with different $k_{\pm}$ and $\omega$
\begin{equation}\label{39}
\Psi = \int e^{-\left(k_{+}^{2}+k_{-}^{2}+\omega^{2}\right)} \psi dk_{+}dk_{-}d\omega.
\end{equation}
We directly compute the norm $||\Psi||$ as
\begin{widetext}
\begin{equation}
\label{40}
%\begin{split}
\small{||\Psi || \equiv
\int  \Psi \Psi^{*} d\beta_{+} d\beta_{-}d\beta_{0}=
\int  d\beta_{+}d\beta_{-}d\beta_{0} \int dk_{+}dk_{-}d\omega \int dk_{+}^{\prime}dk_{-}^{\prime}d\omega^{\prime} e^{-\left(k_{+}^{2}+k_{-}^{2}+\omega^{2}+k_{+}^{\prime 2}+k_{-}^{\prime 2}+\omega^{\prime 2}\right)}
\Psi \left(k_{+},k_{-},\omega\right)\Psi^{*}\left(k_{+}^{\prime},k_{-}^{\prime}, \omega^{\prime}\right).}
%\end{split}
\end{equation}
\end{widetext}

Now we integrate over $\beta_{+},\beta_{0},\beta_{-}$ to obtain Dirac-delta functions which in turn are employed to integrate  over $k_{+}^{\prime},k_{-}^{\prime}, \omega^{\prime}$ to obtain
\begin{equation}
\label{41}
||\Psi ||= \int dk_{+} e^{-2k_{+}^{2}} \int dk_{-} e^{-2k_{-}^{2}} \int d\omega e^{-2\omega^{2}}= \left(\frac{\pi}{2}\right)^{\frac{3}{2}}. 
\end{equation}\\

If we wish to have superposition of negative energy states only, we impose the condition $\omega^{2}>\left(k_{+}^{2}+k_{-}^{2}\right)$. This choice arises from the fact that the super Hamiltonian \eqref{11} is constrained to be zero i.e we have $H_{g}+H_{f}=0$. Thus the Hamiltonian (energy) due to gravity part and the Hamiltonian (energy) due to fluid part must add up to zero. If we assume on physical ground that the energy of the fluid is positive definite, then the energy due to the gravity sector has to be negative since it must compensate for the positive energy of the fluid so as to validate the constraint \eqref{11}. In fact, we can have such states due to the hyperbolicity of the kinetic term of the gravity sector of the Hamiltonian \eqref{hamgrav}. It is noteworthy that there is no instability due to this unboundedness of the Hamiltonian of gravity sector since there is always a compensating positive energy of the fluid so as to ascertain the total Super Hamiltonian equal to zero i.e. 
 $H_{g}+H_{f}=0$. \\

Even with this condition imposed, we can find the norm as follows
\begin{equation}
\label{Norm}
\small{
|| \Psi ||= \int dk_{+}dk_{-}  e^{-2k_{+}^{2}-2k_{-}^{2}} \int_{\sqrt{k_{+}^{2}+k_{-}^{2}}}^{\infty} d\omega e^{-2\omega^{2}}= \frac{\sqrt{2}-1}{8}\pi^{\frac{3}{2}}. }
\end{equation}

Hence, contrary to general belief, we clearly have a time independent and finite normed wave packet.  Although $\langle \beta_{i}\rangle =0$ in this case, use of relevant pre-factors in the calculation of the expectation values indicate that  the second moments, $\langle\beta_{i}^{2}\rangle \neq 0$ and also not same for all $i$'s. So the unitarity is not achieved at the expense of the anisotropy itself or the evolution of the universe. It is noteworthy that one can treat $k_{\pm}, E$ as independent variables without going over to $k_{\pm},\omega$. In that case, we will get a  wavepacket different from \eqref{39}. Nonetheless, the norm of that wavepacket can be shown to be finite and time independent as well. \\

We provide a clear counter-example of the alleged non-unitarity in anisotropic cosmological models and thereby show that nonunitarity is not perhaps a built in pathology for anisotropic models. We shall however, show more examples as follows.\\

{\bf Models with other Equations of State($0 < \alpha <1 $):} \\

In equation \eqref{WDeq}, we have assumed a particular ordering, namely,
\begin{equation}
\label{47}
\mathcal{H}=e^{-3\alpha\beta_{0}}\left[-\frac{1}{24}e^{3(\alpha-1)\beta_{0}}\left(p_{0}^{2}-p_{+}^{2} -p_{-}^{2}\right)+p_{T}\right].
\end{equation}

Quantization would require promoting the variables to operators. There is no reason why the ordering of operators chosen in equation \eqref{WDeq} has a special status. Majumder and Banerjee \cite{barun} showed that the norm of the wave packet becomes asymptotically time independent for a different operator ordering in the case of a Bianchi V model. Here we try a different operator ordering given by

 \begin{widetext}
\begin{equation}
\label{48}
\small{
\mathcal{H}=e^{-3\alpha\beta_{0}}\left[-\frac{1}{24}e^{\frac{3}{2}(\alpha-1)\beta_{0}}\left(p_{0}e^{\frac{3}{2}(\alpha-1)\beta_{0}}p_{0}-p_{+}e^{\frac{3}{2}(\alpha-1)\beta_{0}}p_{+} -p_{-}e^{\frac{3}{2}(\alpha-1)\beta_{0}}p_{-}\right)+p_{T}\right].}
\end{equation}
\end{widetext}

With the standard separation of variables as, $\Psi(\beta_{0},\beta_{+},\beta_{-},T)=\phi(\beta_{0})\psi(\beta_{+},\beta_{-})e^{-\imath E T}$,
the equation for $\phi$ becomes
\begin{equation}
\label{Operatororder}
\small{
\left[e^{\frac{3}{2}(\alpha-1)\beta_{0}}\frac{\partial}{\partial \beta_{0}}e^{\frac{3}{2}(\alpha-1)\beta_{0}}\frac{\partial}{\partial\beta_{0}}+k^{2}e^{3\left(\alpha -1\right)\beta_{0}}-24E\right]\phi =0.}
\end{equation}

where $k^{2}=\left(k_{+}^{2}+k_{-}^{2}\right)$. This is because in the $p_{+},p_{-}$ sector, the factor ordering does not matter and again the solution of this sector will be of the form $e^{\imath k_{+}\beta_{+}}e^{\imath k_{-}\beta_{-}}$. This ordering can also be extended to $\alpha=1$ where the exponent $(1-\alpha)$ vanishes trivially and we get back equation \eqref{phi0}.  \\

For $\alpha\neq 1$ we make a change of variables as
\begin{equation}
\label{51}
\chi = e^{-\frac{3}{2}(\alpha-1)\beta_{0}},
\end{equation}
so that equation \eqref{Operatororder} becomes
\begin{equation}
\label{Inversesquare}
\frac{9}{4}\left(1-\alpha\right)^{2}\frac{d^{2}\phi}{d\chi^{2}}+\frac{k_{+}^{2}+k_{-}^{2}}{\chi^{2}}\phi -24E\phi =0.
\end{equation}
 We define
$\sigma =\frac{4\left(k_{+}^{2}+k_{-}^{2}\right)}{9\left(1-\alpha \right)^{2}}$ and $E^{\prime}=\frac{32}{3\left(1-\alpha \right)^{2}}E$ and equation \eqref{Inversesquare} is now written as
\begin{equation}
\label{Inversesquare2}
\mathcal{H}_{g} \phi =\frac{d^{2}\phi}{d\chi^{2}}+\frac{\sigma}{\chi^{2}}\phi = E^{\prime}\phi.
\end{equation}
We can write \eqref{Inversesquare2} in the form
\begin{equation}
-\frac{d^{2}\phi}{d\chi^{2}}-\frac{\sigma}{\chi^{2}}\phi = -E^{\prime}\phi,
\end{equation} 
which is in fact a well known Schrodinger equation of a particle of $m=1/2$ in an attractive inverse square potential with energy $-E^{\prime}$. This potential is extensively studied in other branches of physics. For a review, we refer to \cite{thesis}.\\

This Hamiltonian may not be self-adjoint to start with, but it has a deficiency index which guarantees existence of a family of self-adjoint extensions \cite{vaughn}. To find deficiency index, we look for eigenfunctions of $\mathcal{H}_{g}$ with an imaginary eigenvalue
\begin{equation}\label{S1}
\mathcal{H}_{g}\phi_{\pm}=\pm \imath \phi_{\pm}
\end{equation}
The general solution to \eqref{S1} is given by Hankel functions as
\begin{equation}\label{S2}
\Phi_{\pm}(\chi)=\sqrt{\chi}\left[A_{\pm}H^{(1)}_{\imath \beta}(\imath \lambda_{\pm} \chi)+B_{\pm}H^{(2)}_{\imath \beta}(\imath \lambda_{\pm} \chi)\right]
\end{equation}
where $\lambda_{\pm} =e^{\pm \imath \frac{\pi}{4}}$ and $\beta=\sqrt{\sigma-\frac{1}{4}}$. But $H^{(2)}_{\imath \beta}(\imath \lambda_{\pm} \chi)$ does not belong to the Hilbert space as it diverges for a large $\chi$. Hence we have
\begin{equation}\label{S3}
\Phi_{\pm}(\chi)=A_{\pm}\sqrt{\chi}H^{(1)}_{\imath \beta}(\imath \lambda_{\pm}\chi)
\end{equation}
Let $n_{\pm}$ be the number of linearly independent solutions for $\mathcal{H}_{g}\Phi_{\pm}=\pm\imath\Phi_{\pm}$, so here we find $n_{\pm}=1$ since we have only one well behaved linear independent solution given by \eqref{S3} for each of the eigenvalues $\pm\imath$. This $n_{+}$ and $n_{-}$ are called the deficiency index. It was shown by Neumann \cite{Neumann} that although the Hamiltonian is not self-adjoint to start with, yet, if $n_{+}=n_{-}$ holds good, it is always possible to have a self-adjoint extension of the same. For an inverse square potential, the method is lucidly described by Essin and Griffiths \cite{griffiths}.\\

The energy eigenvalue equation \eqref{Inversesquare2} can be solved for three different regions, for i) $\sigma > \frac{1}{4}$, ii) $\sigma < \frac{1}{4}$ and iii) the critical case $\sigma = \frac{1}{4}$ and one can find solutions which conserve probability. The details of the calculations can be found in \cite{thesis, griffiths, gupta}. The solution to \eqref{Inversesquare2} for $\sigma > \frac{1}{4}$ is given by Hankel functions of imaginary order 
\begin{equation}\label{S12.73}
\phi_{a}(\chi)=\sqrt{\chi}\left[AH^{(2)}_{\imath \beta}(\lambda \chi)+BH_{\imath \beta}^{(1)}(\lambda \chi)\right]
\end{equation}
where $\beta=\sqrt{\sigma-\frac{1}{4}} \in\mathbb{R}$ and the spectrum is given by $E^{\prime}=-\lambda^{2}$. On the other hand, the solution for $\sigma <\frac{1}{4}$ is given by Hankel functions of real order
\begin{equation}\label{Hankel}
\phi_{b}(\chi) =\sqrt{\chi}\left[AH^{(2)}_{\alpha}(\lambda \chi)+BH_{\alpha}^{(1)}(\lambda \chi)\right]
\end{equation}
where $\alpha=\sqrt{\frac{1}{4}-\sigma}\in \mathbb{R}$ and  $E^{\prime}=-\lambda^{2}$. It is noteworthy, for $\sigma=\frac{1}{4}$, the above two solutions \eqref{S12.73}, \eqref{Hankel} merge as $\beta=\alpha=0$.\\

After the self-adjoint extension as described in \cite{griffiths}, using the asymptotic expression for $\phi_{a}$,  the reflection coefficient $|\frac{B}{A}e^{\pi \beta}|^{2}$ is unity, as required by conservation of probability. Similarly, using the asymptotic expression for $\phi_{b}$, we can find the reflection coefficient as $|\frac{B}{A}|^{2}$, which can be shown to be one as well, after the required extension. Thus, we have a self-adjoint Hamiltonian and the evolution is unitary. \\

In the variable $\chi$ the norm is defined by
\begin{equation}
\label{54.1} 
\langle\phi,\psi\rangle = \int d\chi\ \phi^{*}\psi
\end{equation}
which, in terms of $\eta =e^{\beta_0}$, the average scale factor, can be written as (with the help of equation \eqref{51})
\begin{equation}
\label{54.2}
\langle\phi,\psi\rangle =\frac{3}{2}\left(1-\alpha\right) \int d\eta\ {\eta}^{\frac{ 1-3\alpha}{2}} \phi^{*}\psi .
\end{equation}
Hence the measure is different from that found in \cite{alvarenga3}.\\

Although it may not have any direct relevance to the unitarity of the model, we have checked that the expectation values of quantities like $\beta_{\pm}$ and $\beta_{0}$ are quite regular for $\alpha \neq 1$ also. \\

Questions may be raised regarding the particular operator ordering as there is no favoured ordering apriori. We can in fact do better. We can effect a transformation of variables at classical level as $\chi = \exp\left(\frac{3\left(1-\alpha\right)\beta_{0}}{2}\right)$.

With this substitution, equation \eqref{7} can be recast into 
\begin{equation}\label{A2}
\mathcal{L}_{g}=-\frac{6}{n_{0}}\left[\frac{4\dot{\chi}^{2}}{9\left(1-\alpha\right)^{2}}-\chi^{2}\left(\dot{\beta}_{+}^{2}+\dot{\beta}_{-}^{2}\right)\right],
\end{equation}
where $n=n_{0}e^{3\alpha\beta_{0}}$. This is quite legitimate since one can rescale the lapse function. The corresponding Hamiltonian for the gravity sector will be
\begin{equation}\label{A3}
H_{g}=-\frac{n_{0}}{24}\left[\frac{9\left(1-\alpha\right)^{2}}{4}p_{\chi}^{2}-\frac{p_{+}^{2}}{\chi^{2}}-\frac{p_{-}^{2}}{\chi^{2}}\right],
\end{equation}
and the super Hamiltonian becomes
\begin{equation}
\label{A4}
\mathcal{H}=-\frac{n_{0}}{24}\left[\frac{9\left(1-\alpha\right)^{2}}{4}p_{\chi}^{2}-\frac{p_{+}^{2}}{\chi^{2}}-\frac{p_{-}^{2}}{\chi^{2}}-24p_{T}\right].
\end{equation}
Variation with respect to $n_{0}$ yields constraint equation for Bianchi-I model:
\begin{equation}
\label{A5}
\left[\frac{9\left(1-\alpha\right)^{2}}{4}p_{\chi}^{2}-\frac{p_{+}^{2}}{\chi^{2}}-\frac{p_{-}^{2}}{\chi^{2}}-24p_{T}\right]=0.
\end{equation}
On quantization, we have Wheeler-DeWitt equation for Bianchi-I model as
\begin{equation}\label{A6}
\frac{9\left(1-\alpha\right)^{2}}{4}\frac{\partial^{2}\Psi}{\partial\chi^{2}}-\frac{1}{\chi^{2}}\frac{\partial^{2}\Psi}{\partial\beta_{+}^{2}}-\frac{1}{\chi^{2}}\frac{\partial^{2}\Psi}{\partial\beta_{-}^{2}}=24\imath\frac{\partial\Psi}{\partial T}.
\end{equation}
With the separability ansatz $\Psi = \phi(\chi)e^{\imath\left(k_{+}\beta_{+}+k_{-}\beta_{-}\right)}e^{-\imath ET}$,
we get back equation \eqref{Inversesquare} from \eqref{A6}. \\ 

As the transformation of variables is implemented at the classical level and no particular operator ordering in the quantization scheme has now been resorted to, the process looks completely transparent.\\

The whole point is to emphasize that the non-unitarity as observed in Bianchi-I model by Alvarenga et. al. \cite{alvarenga3} may be due to a bad choice of co-ordinates. Either a clever operator ordering or even a suitable change of variables at the outset would allow us to have a self-adjoint extension of the Hamiltonian and hence one is able to find solutions such that reflection coefficient takes a value so as to conserve the probability and render the model unitary. The detailed method is elegantly discussed by Essin and Griffiths \cite{griffiths}. It also deserves mention that if unitarity is restored, there is a indeed consistency between the expectation values and the Bohmian trajectories as shown in \cite{alvarenga3}.\\

It deserves mention that inverse square potentials may lead to some trouble, it may lead to some strange behaviour, and in many a case the properties, albeit they are useful, are derived only approximately. In a theory of gravitation, the implications have to be thoroughly investigated. For instance, the classical geodesic equation are to be studied carefully to check if there is any pathology like a geodesic incompleteness. For this, more examples will have to be worked out if possible. Anyway, anisotropic quantum cosmological models should now be investigated carefully in order to ascertain the price, if any, one has to pay to secure unitarity.\\

The case for the stiff fluid ($\alpha = 1)$ is quite straight forward. One example is enough to show that the nonunitarity is not generic to anisotropic models. However, we conclusively show that for other equations of state also, we can construct models that conserve probabilty. So the conclusion is that the legendary pathological behaviour of anisotropic quantum cosmological models is not generic. One needs to find either a suitably crafted  operator ordering  or a completely non-controversial but favourable choice of coordinates. Now that one has wave functions which preserve the conservation of probability, the wave functions are worth looking at more closely for various physical aspects, particularly the quantum effects, even for anisotropic models. \\

\begin{acknowledgments}
The authors would thank Golam Hossain and Ritesh Singh for a stimulating discussion.\\
\end{acknowledgments}

\end{document}